# Spherical Layout Implementation using Centroidal Voronoi Tessellations

Martín Larrea, Dana Urribarri, Sergio Martig and Silvia Castro

**Abstract**—The 3D tree visualization faces multiple challenges: the election of an appropriate layout, the use of the interactions that make the data exploration easier and a metaphor that helps in the process of information understanding. A good combination of these elements will result in a visualization that effectively conveys the key features of a complex structure or system to a wide range of users and permits the analytical reasoning process. In previous works we presented the Spherical Layout, a technique for 3D tree visualization that provides an excellent base to achieve those key features. The layout was implemented using the TriSphere algorithm, a method that discretized the spheres's surfaces with triangles to achieve a uniform distribution of the nodes. The goal of this work was centered in a new algorithm for the implementation of the Spherical layout; we called it the Weighted Spherical Centroidal Voronoi Tessellations (WSCVT). In this paper we present a detailed description of this new implementation and a comparison with the TriSphere algorithm.

—————————— ◆ —————————

## 1 INTRODUCTION

INFORMATION Visualization is a very young research field, that has grown very fast as a rich and interdisciplinary research field. The last advances in Visualization, and particularly in Information Visualization, also highlight fundamental research issues. Nowadays, it is currently a challenging task for designers to find out the strategies and tools available to visualize a particular type of information.

The data characteristics and their organization are essential aspects at the adequate visual representation selection. The creation of adequate visual representations is a big challenge. A visual representation is able to convey relationships among many elements in parallel and provides the user with a tool to explore the data in an effective way. Visual representations are essential aids to human cognitive tasks to the extent that they provide stable and external reference points upon which dynamic activities and thought processes may be calibrated and upon which models and theories can be tested and confirmed. The interaction with visual representations makes many complex and intensive cognitive tasks feasible. Visual representations and interaction techniques must allow the users to see, explore and understand large amounts of information at once and are essential to the analytical reasoning process in order to gain insight in the data.

Information Visualization has become a large field and tree visualization has emerged as an important subfield applicable when there is a hierarchical relation among the data elements to be visualized. Tree visualization has many areas of application, and many domains require the manipulation and comprehension of complex hierarchical datasets; to address this field, it is necessary to support interactive representations of large trees. The 3D tree visualization faces multiple challenges: the election of an appropriate layout, the support of the interactions that make the data exploration easier and a metaphor that helps with the information understanding. A good combination of these elements will result in a visualization that effectively conveys the key features of a complex structure or system to a wide range of users and permits the analytical reasoning process.

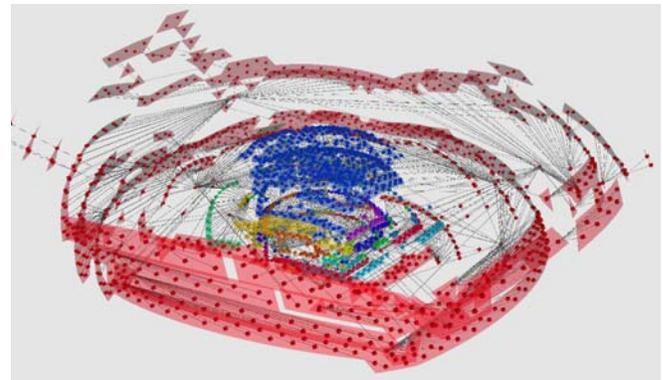

Fig. 1. A file hierarchy visualization using Spherical Layout implemented with the TriSphere algorithm [1].

The Spherical Layout [1], [2] (Fig. 1) is a 3D extension of the 2D radial layout; it was created to allocate a larger number of nodes than the Radial Layout with an intuitive set of interactions in a 3D environment. Although the Spherical Layout does fit a large number of nodes, its distribution in the space, by the TriSphere algorithm, is not

————————————————

- *M.Larrea. D. Urribarri, S. Marting and S. Castro are with the Universidad Nacional del Sur, Departamento de Ciencias e Ingeniería de la Computación, Laboratorio de Investigación y Desarrollo en Visualización y Computación Gráfica. Bahía Blanca, Argentina. CP 8000.*
- *M. Larrea and D. Urribarri are with the Consejo Nacional de Investigaciones Científicas y Técnicas (CONICET), Ciudad de Buenos Aires. CP C1033AAJ.*

optimal. In this paper, we present a new approach for the distribution of the nodes in the space under the Spherical Layout; by using a Spherical Centroidal Voronoi Tessellations (SCVT) we can distribute any number of nodes on the sphere maximizing its surface usage.

This paper is organized as follows: in the next section we present a brief description of the Spherical Layout and its problems. We continue in Section 3 with a description of the Spherical CVT. Afterwards, in Section 4, we compare both algorithms. Section 5 ends this paper with the conclusions and future work.

## 2 SPHERICAL LAYOUT

We have presented previously the Spherical Layout as an extension of the Radial Layout to three dimensions. Such extension is no straightforward, because the nodes must be placed on a surface instead of an arc and this adds many alternatives. In this section we give a brief presentation of the Spherical layout; as we said at the beginning of this paper; a good association of layout and interactions will result in an effective visualization. For the Spherical Layout we had defined a set of interactions, for details of these interactions and more please refer to [1].

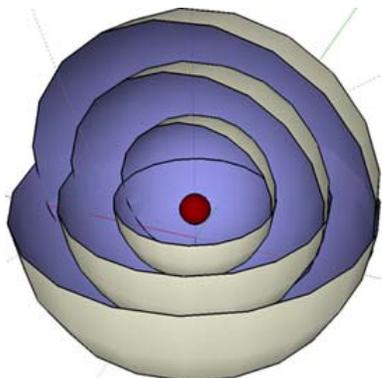

Fig. 2. The concentric spheres are the basis of the Spherical Layout.

The basis of the Radial Layout are the concentric circles where the nodes are placed; the first step to a 3D generalization is to map these circles into a 3D space. To achieve this goal we consider concentric spheres on which surfaces the nodes are going to be placed (Fig. 2). In the Radial Layout each node, except the root, is allocated in a 2D sector within the sector assigned to its parent; in the Spherical Layout we replace the 2D plane with a 3D region and the nodes are allocated within the surfaces defined by it (Fig. 3).

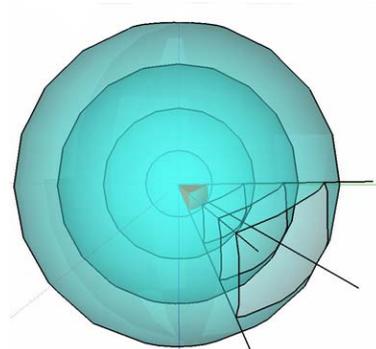

Fig. 3. In Spherical Layout the region assigned to a node is defined by a pyramid.

The implementation presented in [1] and [2] is describe as follow: In this implementation the nodes are uniformly distributed on the spheres surfaces; to achieve this goal we first discretized the surfaces of the spheres with triangles and place the nodes in the center of them. We create as many concentric spheres as levels the tree has, all with the same number of triangles. The Spherical Layout discretized uniformly the surfaces of the spheres with triangles; in order to achieve this, we start with an icosahedron, 20 triangles or faces. If the amount of nodes to be allocated is smaller or equal to 20, we place each node in each of the icosahedron's faces. To allocate more than 20 nodes we increased the number of triangles in our sphere. To maintain a uniform distribution each triangle is divided into four triangles; so the number of triangles increases by a factor of 4. Starting from 20 triangles, the next division will result in 80, then 320 and so on. The hierarchical relations of the tree are present here through the projection of the region and subregion from one sphere to the next inner sphere. The execution time of this algorithm is in the order of the number of leaves by the depth of the tree.

### 2.1 TriSphere Distribution Problem

As we can see in figure 4, if the number of nodes is significantly smaller that 20 the result is not the best distribution. To allocate more than 20 nodes we increase the number of triangles in our sphere. If the number of nodes has the form $20*4^i$, where $i$ is a positive integer, the resulting distribution will be perfect (Fig. 5). However, any number different from this will result in a non optimal distribution.

Let us assume we have to allocate 100 nodes; a sphere with 20 triangles is not enough, neither with 80, we must create a sphere with 320 triangles. From those 320 triangles only 100 will be used, which means that 220, more than 65% of the surface, will not be used.

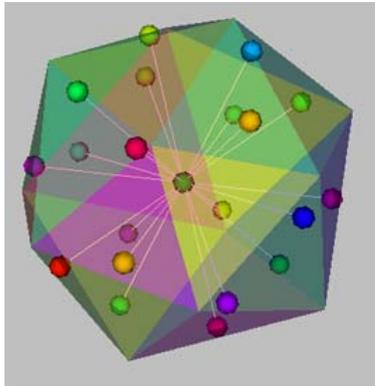

Fig. 4. Twenty nodes distributed on the sphere's surface. Because the number of nodes is equal to the number of triangles in the surface we achieve a perfect distribution.

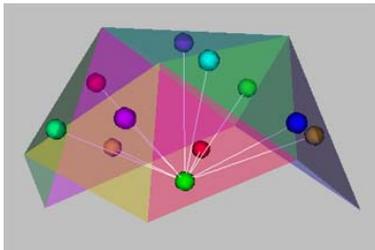

Fig. 5. Ten nodes distributed on the sphere's surface. Because we used an icosahedron as the smallest sphere, 20 triangles, half the surface is not used.

## 3 WEIGHTED SPHERICAL CVT

In order to overcame the distribution problem on the sphere's surface, and based on the Weighted Centroidal Voronoi Tessellation presented in [3] we developed a new algorithm for the node distribution, the Weighted Spherical Centroidal Voronoi Tessellation. Our goal was to make the most of the entire surface of the sphere at the moment of distributing nodes on it. Several work have been done on Spherical Voronoi Tessellation ([9]) and Spherical Centroidal Voronoi Tessellation ([7], [6], [8]), including Constrained Centroidal Voronoi Tessellation, but none of this work refers to Weighted Centroidal Voronoi Tessellation on the surface of the sphere. In contrast to Constrained Centroidal Voronoi Tessellation, which associates a density function to the surface, the Weighted Centroidal Voronoi Tessellation associates values (weights) to the generators: a greater weight means a bigger generated region, independently of the position of the generator on the surface.

Given a set of points (generators) $P = \{p_1, p_2, ..., p_n\}$ in $R^m$, a Voronoi Tessellation is a set of $n$ regions $V(p_i)$, where a point $q \in R^m$ lies in region $V(p_i)$ if and only if $distance(p_i, q) < distance(p_j, q)$ for each $p_i, p_j \in P, i \neq j$.

A Spherical Voronoi Tessellation is a Voronoi Tessellation of the surface of a sphere. In this case the set $P$ is a set of points lying on a surface $S = \{(x, y, z) \in R^3 : x^2 + y^2 + z^2 = 1\}$, and the regions $V(p_i)$ are the points $q \in S$ which satisfy $distance(p_i, q) < distance(p_j, q)$ for each $p_i, p_j \in P, i \neq j$.

A Weighted Spherical Voronoi Tessellation (WSVT) is a Spherical Voronoi Tessellation where each generator $p_i$ has associated a weight $w_i$, and the distance between a point $a$ and a generator $p_i$ is the weighted distance $w-distance(a, w_a, x) = |a - x|^2 - w_a$, where $|.|$ is the euclidean distance.

The general idea of the algorithm used to calculate the WSVT of points $P$ on the sphere is detailed in algorithm 1. The weighted circumcenter of a spherical triangle $\Delta abc$ where $w_a$, $w_b$ and $w_c$ are the weights of $a$, $b$ and $c$ respectively, is defined as follows. Let $x$ be the point coplanar to $a$, $b$ and $c$ which satisfies $w-distance(a, w_a, x) = w-distance(b, w_b, x) = w-distance(c, w_c, x)$. Then, the weighted circumcenter of spherical $\Delta abc$ is $x/|x|$.

**Algorithm 1** Weighted Spherical Voronoi Tessellation (WSVT)

**Input** A set of points $P = \{p_1, ..., p_n\}$ and a set of weights $W = \{w_1, ..., w_n\}$ where $w_i$ is the weight of $p_i$.
**Output** The WSVT $V$ of $P$ and $W$ on the spherical surface $S$.

**1.** Let $H$ be the Convex Hull of $P$ in $R^3$. It represents the Delaunay Triangulation of $P$ on $S$. Note that $H$ is not weighted.
**2.** Let $V$ be the Voronoi Tessellation constructed as the dual graph of $H$: for each triangle in $H$ its weighted circumcenter is a vertex in $V$. If two triangles in $H$ are neighbors then their weighted circumcenters are linked by an edge in $V$.
**3.** Return $V$.

A Centroidal Voronoi Tessellation (CVT) [5] is a particular Voronoi Tessellation where the generator of each Voronoi region is the center of mass (centroid) of its own region.

A CVT with weighted distance is appropriate to divide a surface into subareas where the size of each one depends on the generator itself and not on the generator's position on the surface.

To calculate the Weighted Spherical Centroidal Voronoi Tessellation (WSCVT) it is necessary to introduce the definition of centroid of a spherical triangle and centroid of a spherical polygon. For the next formulae we are considering triangles and polygons on a unitary sphere, therefore the radius is 1 in all of them.

The centroid of a spherical triangle $\Delta abc$ is $(a+b+c)/|a+b+c|$. The centroid of a spherical polygon $v_0, ..., v_n$ [4] is

$$\frac{\sum area(\Delta v_0 v_i v_{i+1}) centroid(\Delta v_0 v_i v_{i+1})}{\sum area(\Delta v_0 v_i v_{i+1})}$$

where the area of triangle $\Delta abc$ is equals to its spherical excess $E$, being $A$, $B$ and $C$ the side lenghts, and $s$ the semiperimeter.

$$E = 4\arctan\sqrt{\tan\left(\frac{1}{2}s\right)\tan\left[\frac{1}{2}(s-A)\right]\tan\left[\frac{1}{2}(s-B)\right]\tan\left[\frac{1}{2}(s-C)\right]}$$

Our WSCVT algorithm is based on the one presented in [3] to compute Weighted CVTs in planar surfaces.

The general idea of the algorithm is to construct the WSVT of a set of generators and then, replace each generator with the centroid of its corresponding Voronoi region, until a desired error has been reached. To control the size of each region, the weighted distance is not enough, it is necesary to adjust the weight of each generator in every iteration: if after one iteration a region size results bigger than the desired size value, the weight of that generator might be decreased for the next iteration, analougsly, if the region size results smaller, the weight might be increased.

It is important to note that the size values are not areas, but percentages. Making an association between the weights of the generators and the surface of the sphere, the area of a region $G_i$ must represent the same percentage of the entire spherical surface as the weight $w_i$ represents of the overall sum of weights.

Then, the desired size value $d_i$ and the actual size value $a_i$ of a region $G_i$ are

$$d_i = \frac{w_i}{\sum w_i} \qquad a_i = \frac{area(G_i)}{area(S)}$$

The algorithm stops when the difference between the desired size value and the actual size value of every region is below a given error $\varepsilon$. The adjusted weight of a generator $p_i$ of current weight $w_i$, desired size value $d_i$ and actual size value $a_i$ is

$$w_i(1 + \frac{a_i - d_i}{d_i})$$

if this value is greater than $\delta$, otherwise the adjusted weight is $\delta$, where $\delta$ is a positive value close to 0, for instance $10^{-6}$, which avoids sites with null weight.

Taking into account the weight adjustment and the size value measure, the algorithm to generate a WSCVT is outlined in algorithm 2. Figure 6 shows the resultant Voronoi Diagram for 100 generators with weights from 1 to 100.

WSCVT can be applied to the first level of the Spherical Layout if the weight of each root's child measures how wide is the representation of the subtree rooted on it. Then each Voronoi region is projected to the outer spheres (Figure 3) to define the pyramid that delimits the regions to place the descendant nodes.

**Weighted Spherical Centroidal Voronoi Tessellation (WSCVT)** *Place nodes on the surface of a sphere*

**Input** A set of weights $W = \{w_1, \ldots, w_n\}$.

**Output** A set of points $P = \{p_1, \ldots, p_n\}$, each point corresponds to the position of a node distributed according $W$.

**1.** Let $P$ be a initial tentative point distribution on a unitary sphere $S$. *(Possibly a random distribution)*

**2.** Let $D$ be the desired values ($d_i = w_i / \sum w_i$)

**3.** while $\varepsilon_{max} > \varepsilon$ *(it has not achieve the desired threshold)*

**4.** Let $V$ be the WSVT of $P$.

**5.** Let $\varepsilon_{actual}$ be the maximum (or average) difference between desired size values ($d_i$) and actual size values ($area(G_i)/area(S)$) of the regions of $V$.

**6.** for all region $G_i$ of $V$

**7.**     Let $p_i \in P$ the generator of $G_i$.

**8.**     Replace $p_i$ with the spherical centroid of $G_i$.

**9.** return $P$.

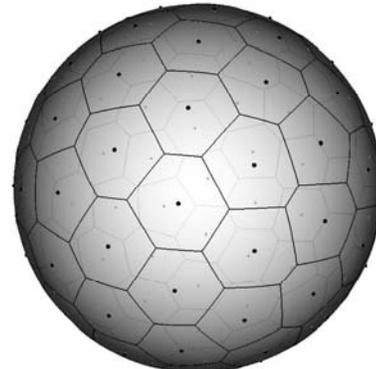

Fig. 6. The Voronoi Diagram of 100 generators, with weights from 1 to 100, with an error less than 5 × 10-4. Transparence has been added to show the back of the sphere.

## 4 TRISPHERE AND WSCVT: A COMPARISON

In this section, we present a brief comparison between the TriSphere and the Weighted Spherical CVT algorithm (Table 1). For each one, we generated 4 trees with different size of; 20, 50, 1000 and 1500 nodes each. In Table 1-(a), because the number of nodes has the form 20*4[i], where $i$ is a positive integer, the resulting distribution using the TriSphere algorithm is perfect; as well as the WSCVT one (Table 1-(b)). In Table 1-(c), in order to allocate 50 nodes using the TriSphere algorithm, the icosahedron must be divided once by four. The resulting figure has 80 faces, which means than 37.5% of the sphere's surface is not used. Table 1-(e) and (g) also shown how the TriSphere algorithm cannot achieve an optimal distribution (21.875% of the surface is not used in the first case and 70.70% in the second one), whereas the WSCVT algorithm always does.

TABLE 1
TRISPHERE AND WSCVT COMPARISON

| Number of nodes in the surface | Algorithm | |
| --- | --- | --- |
| | TRISPHERE | WSCVT |
| 20 | 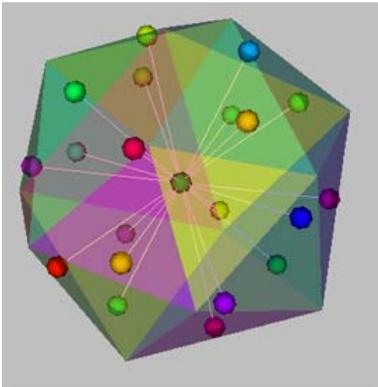<br>(a) 0% of the surface is not used. | 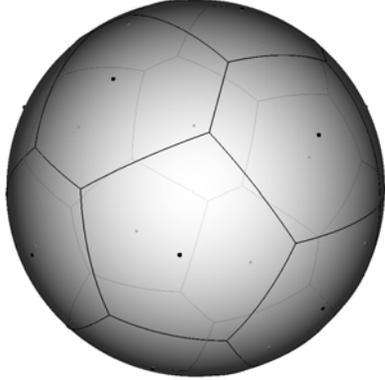<br>(b) 0% of the surface is not used. |
| 50 | 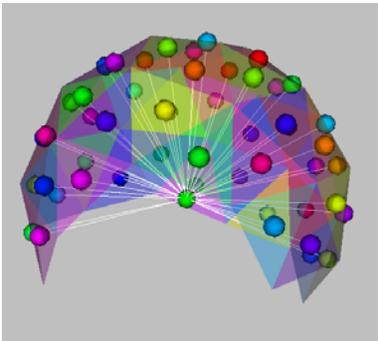<br>(c) 37.5% of the surface is not used. | 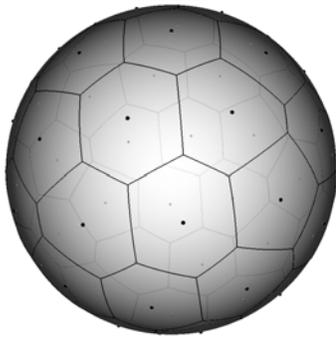<br>(d) 0% of the surface is not used. |
| 1000 | 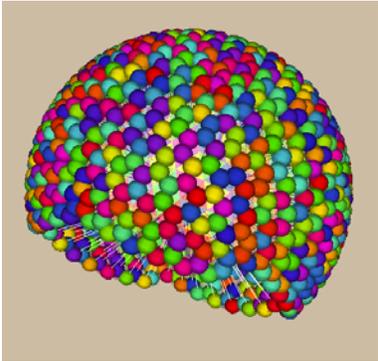<br>(e) More than 21% of the surface is not used | 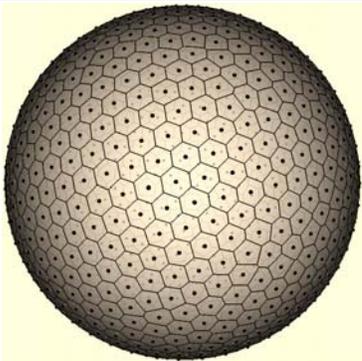<br>(f) 0% of the surface is not used. |
| 1500 | 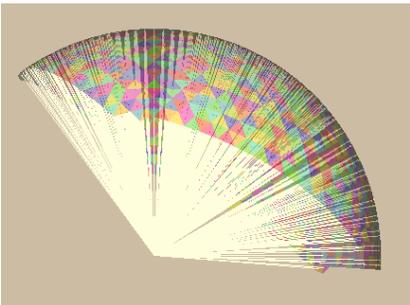<br>(g) More than 70% of the surfaces is not used. | 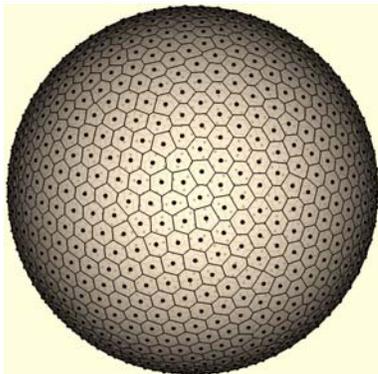<br>(h) 0% of the surface is not used. |

## 5 CONCLUSIONS AND FUTURE WORK

We have developed a new algorithm for the Spherical layout, the WSCVT. Based on the Weighted Centroidal Voronoi Tessellation, WSCVT can make the most of the entire surface of the sphere at the moment of distributing nodes by associating values (weights) to the generators. Unlike the TriSphere algorithm, WSCVT always used the entire sphere's surface to position the nodes.

WSCVT algorithm can still be highly improved. Due to the weighted part of the algorithm being based on a non-weighted one, some wrong edges may exist, and then some overlapping regions may appear, especially when the differences between the weights are considerable (the weight set has a high variance).

One easy solution, but not always effective, is to traverse the edges of the Delaunay Triangulation, and swap every wrong edge. This procedure reduces the existance of wrong edges, althought it cannot always eliminate all wrong edges.

This is a time consuming solution to the problem and not an adequate one, therefore we are working on designing a complete weighted algorithm. Furthermore, we are looking into the possibility of applying this algorithm to just a portion of the surface.

### ACKNOWLEDGMENT

This work was partially supported by the PGI 24/N020, 24/ZN12 and 24/ZN19, Secretaría General de Ciencia y Tecnología, Universidad Nacional del Sur, Bahía Blanca, Argentina.

**Martín L. Larrea** is a member of the Computer Graphics and Visualization R&D Laboratory (VyGLab) at the Universidad Nacional del Sur. Larrea received his degree in Computer Science in 2004 and a Master in Computer Science in 2006, both from the Universidad Nacional del Sur. Martín Larrea is member of three ongoing research projects funded by the argentinean government. He is currently finishing his PhD on Semantic Based Visualization

**Dana K. Urribarri** is a member of the Computer Graphics and Visualization R&D Laboratory (VyGLab). Urribarri received her degree in Computer Engineering in 2006.

**Sergio R. Martig** is a member of the Computer Graphics and Visualization R&D Laboratory (VyGLab).

**Silvia M. Castro** is a member of the Computer Graphics and Visualization R&D Laboratory (VyGLab) and of the ACM Computer Society.